\documentclass[showpacs,pre]{revtex4}
\usepackage{graphicx}
\begin{document}
\title{Efficiency and Fluctuation in Tight-Coupling Model of Molecular Motor}
\author{Hidetsugu Sakaguchi}
\address{Department of Applied Science for Electronics and Materials,\\
Interdisciplinary Graduate School of Engineering Sciences,\\
Kyushu University, Kasuga, Fukuoka 816-8580, Japan}
\begin{abstract}
A simple tight-coupling model of a molecular chemical engine is proposed. 
The efficiency of the chemical engine and its average velocity can be explicitly calculated.  The diffusion constant is evaluated approximately using the fluctuation theorem. Langevin simulations with stochastic boundary conditions are performed and the numerical results are compared with theoretical calculations.
\end{abstract}
\maketitle

Molecular motors such as myosin, kinesin, and F$_1$-ATPase are highly efficient for converting chemical energy into mechanical work, although the mechanism of the energy transduction is not fully understood.~\cite{rf:1,rf:2,rf:3,rf:4}
Since the molecular motors are very small, they work under strong thermal fluctuations. Various ratchet models have been intensively studied, in which the fluctuations play an important role for the directional motion.~\cite{rf:5,rf:6,rf:7}  
In particular, F$_1$-ATPase is a reversible chemical engine, in which the torque and the rotary motion are induced by the chemical reaction: ATP$\rightarrow$ ADP$+$P, and ATP is produced from ADP$+$P by the reverse torque. It is known that the efficiency of the energy conversion is close to 1.~\cite{rf:4} 
Some ratchet-type models were proposed for the rotary chemical engine.~\cite{rf:8,rf:9,rf:10}
We proposed a simple Langevin model with stochastic boundary conditions for the Feynman ratchet~\cite{rf:11} and showed that entropy production satisfies the fluctuation theorem.~\cite{rf:12}
In this letter, we will generalize the Langevin model with stochastic boundary conditions for a chemical engine such as the F$_1$-ATPase. The model is simple such that the average rotational velocity and the efficiency of energy conversion can be obtained explicitly; further, the diffusion constant for the Brownian motion is approximately evaluated. 

We propose a simple model of a rotary molecular motor such as F$_1$-ATPase under thermal fluctuation at temperature $T$. 
The angle of the rotary machine is denoted as $x$.  The average velocity of the Brownian motion of the molecular machine is assumed to be $v_1$ in the region    $x<x_1$ and $x>x_2$ and $v_2$ in the region $x_1<x<x_2$. Here, we have assumed two special angles $x_1$ and $x_2$. The chemical energy of ATP serves as the energy source of the molecular motor. ATP is adsorbed only at the angle $x_1$.  If chemical energy is provided to the motor, the motor can rotate further. 
A similar type of local reaction was also assumed in ref.~10. 
The adsorption of ATP is assumed to occur in a stochastic manner. We introduce a probability $p_1$, which is proportional to the adsorption probability of ATP at $x=x_1$.  The Brownian particle can pass through $x_1$ in the $+x$-direction with the passing probability $p_1$, although the Brownian particle can pass through $x_1$ freely in the $-x$-direction. This type of stochastic boundary condition was used in a model for the Feynman ratchet.~\cite{rf:11}
If ATP adsorption does not occur, the Brownian particle stays within the region $x<x_1$. The adsorbed ATP changes into ADP$+$P in the region $x_1<x<x_2$, and ADP is desorbed at another specific angle $x_2$. The Brownian particle can pass through $x_2$ freely in the $+x$ direction, and the reverse motion occurs only with a probability $p_2$, which is proportional to the adsorption probability of ADP at $x=x_2$.  The average velocity after the release of ADP is again $v_1$. 

The Langevin equation for $x$ is given by 
\begin{eqnarray}
\frac{dx}{dt}&=&v_1+\xi(t), \;\;\;{\rm for }\;\; 0<x<x_1,\;x_2<x<x_0,\nonumber\\
\frac{dx}{dt}&=&v_2+\xi(t), \;\;\;{\rm for }\;\; x_1<x<x_2,
\end{eqnarray}
where $\xi(t)$ denotes the Gaussian white noise satisfying $\langle \xi(t)\xi(t^{\prime})\rangle=2T\delta(t-t^{\prime})$, and the spatial period of $x_0=2\pi/3$ is assumed, because F$_1$-ATPase has a rotational symmetry of the angle $2\pi/3$.  
Equation (1) can be rewritten using a potential as 
\begin{equation}
\frac{dx}{dt}=-\frac{\partial U(x)}{\partial x}+\xi(t),
\end{equation}
where  $U(x)=-v_1x$ for $0<x<x_1$, $U(x)=-v_2 (x-x_2)+U_0$ for $x_1<x<x_2$, and $U(x)=-v_1(x-x_0)$ for $x_2<x<x_0$. The parameter values of $x_1=\pi/9$ and $x_2=5\pi/9$ are used in this paper.
The probabilities $p_1$ and $p_2$ are assumed to be $p_1=\exp(-\Delta E/T)\exp(-\Delta \mu/T)\exp\{(-v_1l_1-v_2l_2)/T\}$ and   
$p_2=\exp(-\Delta E/T)$, where $l_1=x_0+x_1-x_2$ and $l_2=x_2-x_1$.
\begin{figure}[htb]
\begin{center}
\includegraphics[width=12cm]{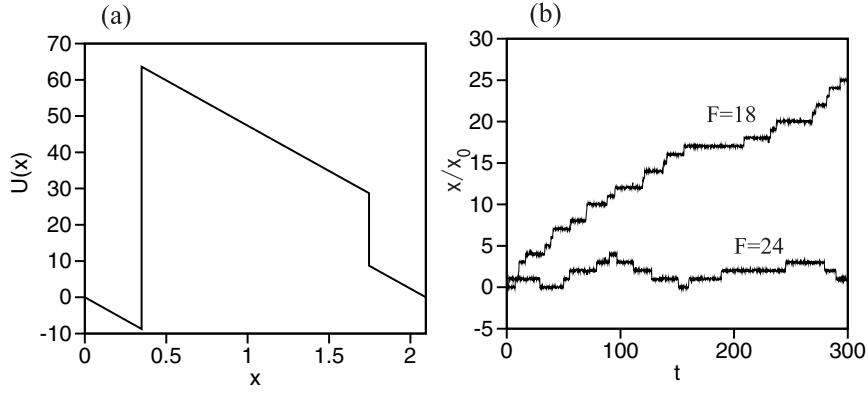}
\end{center}
\caption{(a) Potential $U(x)$ for $v_1=v_2=25$ and $\Delta E=20$. (b) Time evolutions of $x(t)$ by Eq.~(3) at $F=18$ and $F=24$ for $v_1=v_2=25$, $\Delta E=20$, $\Delta\mu=50$, and $T=4$}
\label{fig:1}
\end{figure}
Figure 1(a) displays the potential $U(x)$ for  $v_1=v_2=25$ and $\Delta E=20$, and the constant $U_0$  is assumed to be $U_0=\Delta E+v_1(x_0-x_2)$.  
The unit of energy is assumed to be pN$\cdot$nm, and the unit of force is pN, although our parameter values are not entirely realistic.  
The potential gap $\Delta U(x_2)$ at $x=x_2$ is then $\Delta E$, and the potential gap $\Delta U(x_1)$ at $x=x_1$ is $\Delta E+v_1l_1+v_2l_2$.
The probabilities $p_1$ and $p_2$ are related to the potential gaps as follows $p_1=\exp(-\Delta U(x_1)/T)\exp(-\Delta \mu/T)$ and $p_2=\exp(-\Delta U(x_2)/T)$.
The energy gap $\Delta U(x_2)=\Delta E$ can be considered as the activation energy  for the adsorption of ADP. The energy gap $\Delta U(x_1)$ is considered as the activation energy for the adsorption of ATP at $x=x_1$. The quantity $\Delta \mu$ denotes a difference in the chemical potential of ATP and ADP$+$P. If ATP and ADP$+$P are in a state of chemical equilibrium, the difference in their chemical potentials $\Delta \mu$ is zero, and the total system comprising ATP,ADP, P, and the molecular machine is in a state of thermal equilibrium. If the concentration of ATP is greater than that in  chemical equilibrium, $\Delta\mu$ becomes negative, and the adsorption probability $p_1$ increases; the molecular motor then tends to rotate in the $+x$-direction. The difference in the chemical potential is expected to obey
\[\Delta \mu=\Delta\mu_0+T\ln[ADP][P]/[ATP]\]  
for an ideal solution. In this case, the adsorption probability $p_1\propto \exp(-\Delta\mu/T)$  at $x=x_1$ is proportional to the concentration $[ATP]$ of ATP.   

When the molecular motor works against
the external torque $F$, the model equation (1) is replaced with 
\begin{eqnarray}
\frac{dx}{dt}&=&v_1-F+\xi(t), \;\;\;{\rm for }\;\; 0<x<x_1,\;x_2<x<x_0,\nonumber\\
\frac{dx}{dt}&=&v_2-F+\xi(t), \;\;\;{\rm for }\;\; x_1<x<x_2.
\end{eqnarray}
In this loaded system, if $x$ is increased from $x=0$ to $x=x_0$, the potential increases from $U(0)$ to $U(x_0)=U(0)+Fx_0$.  When $Fx_0$ becomes equal to the difference in the chemical potential $-\Delta\mu$, the chemical energy of ATP is completely used for the work of the uphill motion. This corresponds to the thermal equilibrium condition. The equilibrium condition does not depend on $\Delta E$, $v_1$, and $v_2$

We have performed numerical simulations of the Langevin equation (3) with the stochastic boundary conditions. The Euler method with a time step of $5\times 10^{-6}$ was used for the numerical simulations. 
Figure 1(b) displays the two time evolutions of $x(t)/x_0$ at $F=18$ and $24$ for $v_1=v_2=25$, $T=4$, $\Delta E=20$, and $\Delta\mu=-50$.
Hereafter, the parameter values of $T=4$, $\Delta E=20$ and $\Delta\mu=-50$ are fixed. 
The Brownian motion exhibits a stepwise motion because the Brownian particle remains within the region $nx_0<x<nx_0+x_1$ or $nx_0+x_2<x<(n+1)x_0$ ($n$ is an integer) for a long duration and occasionally jumps from $nx_0+x_1$ to $nx_0+x_2$ in a relatively short time.  The average velocity is definitely positive for $F=18$, however, it is nearly 0 for $F=24$ because $F=24$ is close to the equilibrium condition $F=F_c=-\Delta\mu/x_0=23.873$. 

\begin{figure}[htb]
\begin{center}
\includegraphics[width=10cm]{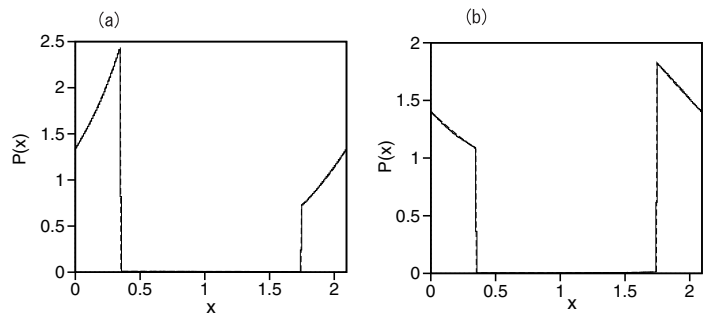}
\end{center}
\caption{Stationary probability distribution function $P(x)$ at $F=18$ and $T=4$ for (a) $v_1=v_2=25$ and (b) $v_1=15$ and $v_2=30$. The numerical results are shown by the solid curves and the theoretical lines are shown by the dashed curves.}
\label{fig:2}
\end{figure}
The corresponding Fokker-Planck equation for the probability density function $P(x,t)$ is given as
\begin{equation}
\frac{\partial P}{\partial t}=-\frac{\partial\{ (-\partial U(x)/\partial x-F)P\}}{\partial x}+T\frac{\partial^2P}{\partial x^2}.
\end{equation}
The stationary probability distribution satisfying the stochastic boundary conditions: $P(x_1-0)\cdot p_1=P(x_1+0)$ and $P(x_2-0)=P(x_2+0)\cdot p_2$ is given by 
\begin{eqnarray}
P(x)&=&\left (P(x_2+0)+\frac{c}{v_1^{\prime}}\right )\exp\{v_1^{\prime}(x+x_0-x_2)/T\}-\frac{c}{v_1^{\prime}},\;\;{\rm for }\;\;0<x<x_1,\nonumber\\
&=&\left (P(x_1+0)+\frac{c}{v_2^{\prime}}\right)\exp\{v_2^{\prime}(x-x_1)/T\}-\frac{c}{v_2^{\prime}},\;\;{\rm for }\;\;x_1<x<x_2,\nonumber\\
 &=&\left(P(x_2+0)+\frac{c}{v_1^{\prime}}\right)\exp\{v_1^{\prime}(x-x_2)/T\}-\frac{c}{v_1^{\prime}},\;\;{\rm for }\;\;x_2<x<x_0,
\end{eqnarray}
where $v_1^{\prime}=v_1-F$, $v_2^{\prime}=v_2-F$, and the three unknown constants $P(x_1+0)$, $P(x_2+0)$, and $c$ are determined by the stochastic boundary conditions and the normalization condition $\int_0^{x_0}P(x)dx=1$.
The constant $c$ is proportional to the probability flow, and $c$ becomes 0 in the equilibrium state.  The average velocity $\langle v\rangle$ of the Brownian motion is given by $-c\cdot x_0$. 
 The explicit form of $c$ is rather complicated, but it is explicitly given by $c=1/(c_1+c_2)$ where 
\[c_1=\frac{\{1-\exp(v_1^{\prime}l_1/T)\}/v_1^{\prime}+\exp(\Delta E/T)\{\exp(v_1^{\prime}l_1/T)-\exp(v_2^{\prime}l_2/T+v_1^{\prime}l_1/T)\}/v_2^{\prime}}{\exp\{(-\Delta\mu-Fx_0)/T\}-1}\]
\[\times \, T[\{1-\exp(-v_1^{\prime}l_1/T)\}/v_1^{\prime}+\exp(-\Delta\mu/T)\exp\{(-\Delta E-v_1l_1-v_2l_2)/T\}\{\exp(v_2^{\prime}l_2/T)-1\}/v_2^{\prime},\}]\]
\[c_2=T\{1-\exp(-v_1^{\prime}l_1/T)\}/v_1^{\prime 2}-l_1/v_1^{\prime}+T\{\exp(v_2^{\prime}l_2/T)-1\}/v_2^{\prime 2}-l_2/v_2^{\prime}.\]

Figure 2(a) and 2(b) display the stationary distribution of $P(x)$ for (a) $v_1=v_2=25$ and (b) $v_1=15$ and $v_2=30$ at $F=18$. 
The solid lines represent the histograms obtained from the Langevin simulations and the dashed curves represent eq.~(5).  The probability in the region $x_1<x<x_2$ is very small. This is because the passing probabilities $p_1$ and $p_2$ are rather small, which is related to the stepwise motion shown in Fig.~1(b).  The probability distribution function $P(x)$ is an increasing function of $x$ in the region $x<x_1$ and $x>x_2$ for $v_1=v_2=25$ because $v_1>F$. On the other hand, $P(x)$ is a decreasing function of $x$ for $v_1=15$ and $v_2=30$ because  $v_1<F$. However, the average velocity of the Brownian motion is positive in both the cases. 
\begin{figure}[htb]
\begin{center}
\includegraphics[width=10cm]{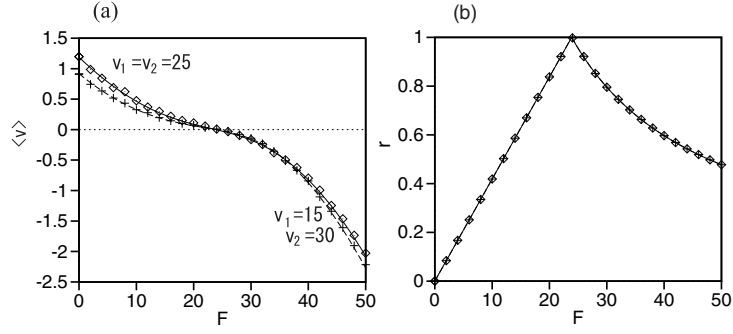}
\end{center}
\caption{(a) Average velocity $\langle v\rangle$ of the rotary Brownian motion as a function of $F$ for $v_1=v_2=25$ (denoted by rhombi) and for $v_1=15$ and $v_2=30$ (denoted by crosses). The theoretical values are shown by the solid and dashed curves. (b) Efficiency $r$ of the rotary molecular motor as a function of $F$ for $v_1=v_2=25$ and for $v_1=15$ (denoted by rhombi) and $v_2=30$ (denoted by crosses).}
\label{fig:3}
\end{figure}

Figure 3(a) displays the average velocity $\langle v\rangle$ in two cases of $v_1=v_2=25$, and  $v_1=15, v_2=30$ as a function of $F$. The symbols represent numerical results of the Langevin simulations, and the curves represent the theoretical values. 
The average velocity becomes 0 at $F=F_c=50/x_0$ in both the cases, however, the average velocities take different values for $F\ne F_c$. The average velocity is a nonlinear function of $F$. 
The efficiency $r$ of the energy conversion is defined as the ratio of the chemical potential and the work per cycle: $r=Fx_0/(-\Delta \mu)$ for $\langle v\rangle>0$. If the average velocity is negative, the external torque induces a reverse rotation. In the case of the reverse rotation, when $x$ decreases from $x=x_2$ to $x=x_1$, ADP$+$P changes to ATP. This is because our molecular motor is a tight-coupling model, and the mechanical work is used for the production of ATP. The efficiency of the ATP production is defined as $r=(-\Delta \mu)/Fx_0$ for $\langle v\rangle<0$.  Figure 3(b) represents $r$ as a function of $F$.  The efficiency attains the maximum value of $r=1$ at $F=F_c$. The symbols denote numerical results for the two cases $v_1=v_2=25$ and $v_1=15,\,v_2=30$. In the numerical simulations, $r$ is calculated as the long time average of the ratio of $Fx(t)$ to $-\Delta \mu\times N_r$, where $N_r$ is the total rotation number of the Brownian motion.

\begin{figure}[htb]
\begin{center}
\includegraphics[width=10cm]{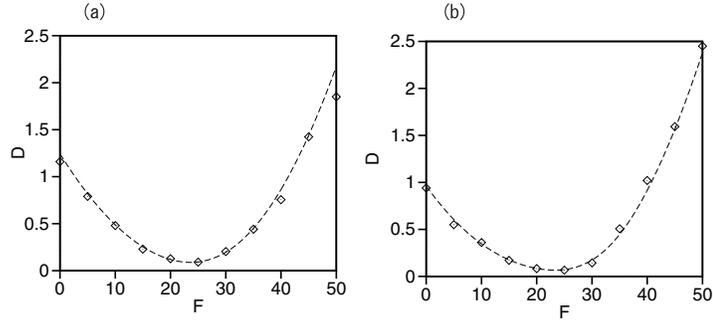}
\end{center}
\caption{Diffusion constant $D$ of the Brownian motion as a function of $F$ for (a) $v_1=v_2=25$ and (b) $v_1=15$, $v_2=30$. The dashed curves are the lines generated by using eq.~(7).}
\label{fig:4}
\end{figure}
The fluctuation of the molecular motor as shown in Fig.~1(b) can be characterized by the diffusion constant $D$ of the Brownian motion. It is defined as \[D=\langle (x(t+t_0)-x(t_0)-\langle x(t+t_0)-x(t_0)\rangle )^2\rangle/(2 t),\]
where the average is taken as the ensemble average. The diffusion constant of the Brownian motion in tilted periodic potentials was studied in refs.~13 and 14.
Here, we evaluate the diffusion constant using the fluctuation theorem.~\cite{rf:15}  The fluctuation theorem is satisfied by a rather large class of stochastic systems, including our model. The probability $P(\sigma)$ of the entropy production $\sigma(\tau)$ during a large time interval $\tau$  satisfies the relation\[\ln \frac{P(\sigma)}{P(-\sigma)}=\sigma(\tau).\]

Because the Brownian motion exhibits a stepwise motion, we consider a spatially@discretized lattice system with lattice constant $x_0$.  The Brownian motion is approximated by a random walk on the lattice.  
We can consider that the Brownian particle stays at the $n$th lattice site with probability $P_n$ defined as $P_n=\int_{nx_0}^{(n+1)x_0}P(x,t)dx$, where $P(x,t)$ is the continuous probability distribution function. 
The time is also discretized as $t_N=N\Delta t$ with the time step $\Delta t$, which is assumed to be sufficiently small. The transition probability from the $n$th site to the $n\pm 1$th site in the short time $\Delta t$ is defined as $p_{\pm}$. The probability that the Brownian particle stays at the $n$th site is given by $p_0=1-p_{+}-p_{-}$. This random walk problem can be explicitly solved. The probability that the number of forward steps is $N_{+}$,  the number of backward steps is  $N_{-}$, and the number of times that the random walker stays at the same site is $N_0=N-N_{+}-N_{-}$ in the total steps $N$ of the random walk is given by $P(N_{+})=N!/(N_{+}!N_{-}!N_0!)p_{+}^{N_{+}}p_{-}^{N_{-}}p_0^{N_0}$. The entropy production in this process is expressed as $\sigma=(-\Delta\mu-Fx_0)(N_{+}-N_{-})/T$.  Therefore, $P(\sigma)=N!/(N_{+}!N_{-}!N_0!)p_{+}^{N_{+}}p_{-}^{N_{-}}p_0^{N_0}$, and the probability of the reverse process is expressed as $P(-\sigma)=N!/(N_{+}!N_{-}!N_0!)p_{+}^{N_{-}}p_{-}^{N_{+}}p_0^{N_0}$. The value $\ln P(\sigma)/P(-\sigma)$ is calculated as $(N_{+}-N_{-})\ln p_{+}/p_{-}$. Only if $p_{+}/p_{-}=e^{(-\Delta\mu-Fx_0)/T}$, the fluctuation theorem is exactly satisfied even for a small step number $N$. Therefore, the ratio $p_{+}/p_{-}$ needs to satisfy $p_{+}/p_{-}=e^{(-\Delta\mu-Fx_0)/T}$.

The average velocity and the diffusion constant of this random walk is therefore evaluated as \begin{eqnarray}
\langle v\rangle&=&x_0(p_{+}-p_{-})/\Delta t,\nonumber\\
D&=&x_0^2\{(p_{+}+p_{-})-(p_{+}-p_{-})^2\}/(2\Delta t).
\end{eqnarray}
For a sufficiently small $\Delta t$, $p_{+}$ and $p_{-}$ are also sufficiently small; therefore, the term $(p_{+}-p_{-})^2$ in the expression of $D$ is negligible. In the limit of continuous time  $\Delta t\rightarrow 0$, $D$ is expressed as\begin{equation}
D\sim \frac{x_0^2(p_{+}+p_{-})}{2\Delta t}=\frac{x_0\langle v\rangle (p_{+}+p_{-})}{2(p_{+}-p_{-})}=\frac{x_0v}{2{\rm tanh}\{(-\Delta\mu-Fx_0)/(2T)\}}.
\end{equation}

Near $F=F_c=-\Delta \mu/x_0$, $v$ is written as $v=\gamma (F_c-F)$, where $\gamma=-\partial v/\partial F$ at $F=F_c$. The diffusion constant $D$ is then expressed as $D=\gamma\cdot T$. This is the Einstein relation at the equilibrium condition. The Einstein relation in which the diffusion constant is expressed by the response coefficient $\gamma$ is satisfied only near $F=F_c$. 
Equation (7) is interpreted as a generalized relation between the diffusion constant $D$ and the generalized response function $\gamma(F)=v/(F_c-F)$. 
Figure 4 shows the numerically obtained values of $D$ and the relation of eq.~(7) as a function of $F$ for (a) $v_1=v_2=25$ and (b) $v_1=15$, $v_2=30$.  Equation (7) is a fairly good approximation of the diffusion constant. 
This equation implies that an effective temperature can be expressed as $T_{eff}=D/\gamma(F)=\nu/{\rm tanh}(\nu/T)$, where $\nu=x_0(F_c-F)/2$. This is a simple relation, although it is based on an approximation by the random walk. The effective temperature does not depend on detailed parameters such as $v_1$, $v_2$ and so on. The effective temperature $T_{eff}$ is greater than $T$, and the equality is satisfied only at $F=F_c$, because $T_{eff}$ is expressed as $T_{eff}=\{(\nu/T)/{\rm tanh}(\nu/T)\}\cdot T$ and $z/{\rm tanh}z\ge 1$ for any $z$.   

To summarize, we have proposed a simple tight-coupling model for a rotary molecular motor. Because it is a piecewise linear and one-dimensional model, various quantities can be calculated explicitly.  Its efficiency is determined only by the ratio of the difference in the chemical potential $-\Delta \mu$ and the work $F\cdot x_0$. We have calculated the diffusion constant approximately by using the fluctuation theorem; and the numerical results obtained are fairly well consistent with the approximation.

\end{document}